\documentclass[aps,prl,twocolumn,showpacs,superscriptaddress]{revtex4-1}
\usepackage{graphicx}
\usepackage{amsmath,amssymb}
\usepackage[utf8]{inputenc}
\usepackage{natbib}
\usepackage{hyperref}
\usepackage{color}
\hypersetup{colorlinks=true,citecolor={blue},linkcolor={blue},urlcolor={blue}}
\usepackage{units}
\usepackage[english]{babel}
\usepackage[normalem]{ulem}

\begin{document}
\definecolor{orange}{RGB}{255, 69, 0}
\title{Optical bistability under non-resonant excitation in spinor polariton condensates}

\date{\today}

\author{L. Pickup}
\affiliation{School of Physics and Astronomy, University of Southampton, Southampton, SO17
 1BJ, United Kingdom}

\author{K. Kalinin}
\affiliation{Skolkovo Institute of Science and Technology, Skolkovo Innovation Center, Building 3, Moscow 143026, Russian Federation}

\author{A. Askitopoulos}
\affiliation{School of Physics and Astronomy, University of Southampton, Southampton, SO17
 1BJ, United Kingdom}

\author{Z. Hatzopoulos}
\affiliation{Microelectronics Research Group, IESL-FORTH, P.O. Box 1527, 71110 Heraklion, Crete, Greece}
\affiliation{Department of Physics, University of Crete, 71003 Heraklion, Crete, Greece}

\author{P. G. Savvidis}
\affiliation{Microelectronics Research Group, IESL-FORTH, P.O. Box 1527, 71110 Heraklion, Crete, Greece}
\affiliation{Department of Materials Science and Technology, University of Crete, Crete, Greece}
\affiliation{ITMO University, St. Petersburg 197101, Russia}

\author{N. G. Berloff}
\affiliation{Skolkovo Institute of Science and Technology, Skolkovo Innovation Center, Building 3, Moscow 143026, Russian Federation}
\affiliation{Department of Applied Mathematics and Theoretical Physics,
	University of Cambridge, Cambridge CB3 0WA, United Kingdom}

\author{P.G. Lagoudakis}
\affiliation{School of Physics and Astronomy, University of Southampton, Southampton, SO17
 1BJ, United Kingdom}
\affiliation{Skolkovo Institute of Science and Technology, Skolkovo Innovation Center, Building 3, Moscow 143026, Russian Federation}

\begin{abstract}

We realise bistability in the spinor of polariton condensates under non-resonant optical excitation and in the absence of biasing external fields. Numerical modelling of the system using the Ginzburg-Landau equation with an internal Josephson coupling between the two spin components of the condensate qualitatively describes the experimental observations. We demonstrate that polariton spin bistability persists for sweep times in the range of $[10 \mu sec,1  sec]$ offering a promising route to spin switches and spin memory elements.

\end{abstract}

\maketitle
Optical bistability is the phenomenon of an optical system supporting two or more stable states for a given range of driving conditions \cite{abraham_opticalbistabilityReview_1982,gibbs_optical_2012_controllingLightwithLight}. A bistable behavior demonstrates an internal memory of the system, which could potentially be harnessed to form optical transistors and memory elements. It has been observed in systems such as cold atoms \cite{ColdAtomBistability1}, lasers \cite{jung_scaling_1990}, self-electro-optic effect devices (SEED) \cite{forsmann_nonresonant_SEED_1987} and Fabry-P\'{e}rot cavities containing nonlinear materials \cite{{gibbs1976FabryPerotBistability},{smith1977bistableFabryPerot}}. Optical bistability in microcavity polaritons, the bosonic quasi-particles formed by the strong coupling of cavity photons and excitons, was previously demonstrated for resonant/quasi-resonant optical excitation \cite{{Baas1},{Baas2},{SpinMultiNature},{Multistability_Early_2013},{Decoh_bistab_Feb2017},{rodriguez2016dynamic}}, electrical biasing \cite{Ohadi_bistab,Bajoni_Bistability_GaAs_photodiode} and non-resonant electrical injection \cite{M_Amthor_Electrical_Bistability}. In resonantly pumped microcavity polaritons, bistability was described by a Kerr-like nonlinearity resulting from polariton-polariton interactions \cite{Baas1} and by employing an analogy with optical parametric oscillators \cite{Baas2}. Under electrical injection, bistability was observed in the photoluminescence intensity, in the presence of an external magnetic field, and was attributed to the electrostatic screening of the injected charge carriers forming a positive feedback for the backward sweep of the driving current \cite{M_Amthor_Electrical_Bistability}.

Bistability has also been realised in the spinor of a polariton state utilising polaritons' well defined spin and its one-to-one correspondence with the circular polarization of light. Optical control of the polariton spin was employed in the realisation of  spin-bistability and multi-stability, using quasi-resonant optical excitation of a cylindrical mesa \cite{SpinMultiNature} by rotating the polarization of the optical pump. Recently, spin-bistability was also shown under non-resonant optical excitation in a mesa structured sample using a biasing electrical field to induce an energy splitting between two linear polarization modes, while keeping the optical excitation constant \cite{Ohadi_bistab}. Under non-resonant pumping and in the absence of external fields, the preferable route for the operation of all-optical memory elements in polariton circuits, both polariton bistability and spin-bistability remain elusive, despite theoretical predictions \cite{kyriienko_bistability_2014}.

For the observation of polariton bistability or spin-bistability, it is essential that the population fluctuations of polaritons or the relative spin populations, remain sufficiently small so as not to bridge the two stable solutions of the system that would collapse the hysteresis. Polaritons are an inherently open-dissipative, non-equilibrium system that requires continuous pumping to reach steady-state due to the finite cavity lifetime.  In the absence of external electric or magnetic fields, polariton bistability and spin-bistability has been observed only under resonant excitation; the fundamental difference between resonant and non-resonant optical pumping is the presence, in the latter case, of a hot-exciton reservoir, and the concomitant exciton-polariton interactions. Exciton-polariton pair scattering events contribute to the decoherence of polaritons \cite{{Krizhanovski},{askitopoulos_polariton_2013}}, however, the observed bistabilities in polariton systems occur at time scales that can exceed the coherence time of a polariton condensate by a factor of $10^{9}$, suggesting that it is not the exciton induced decoherence that precludes the observation of bistability under non-resonant pumping. Another source of instability that prevents bistable behavior under non-resonant pumping are the density fluctuations induced by the hot-exciton reservoir. Similarly, exciton-polariton interactions are a strong source of spin mixing \cite{{Kavokin_Lagoudakis_PRB}} that, in the presence of an initial spin imbalance, result in changes of the relative spin populations that prevent the realisation of spinor polariton bistability.

\begin{figure}[t!]
\center
\includegraphics[width=8.6cm]{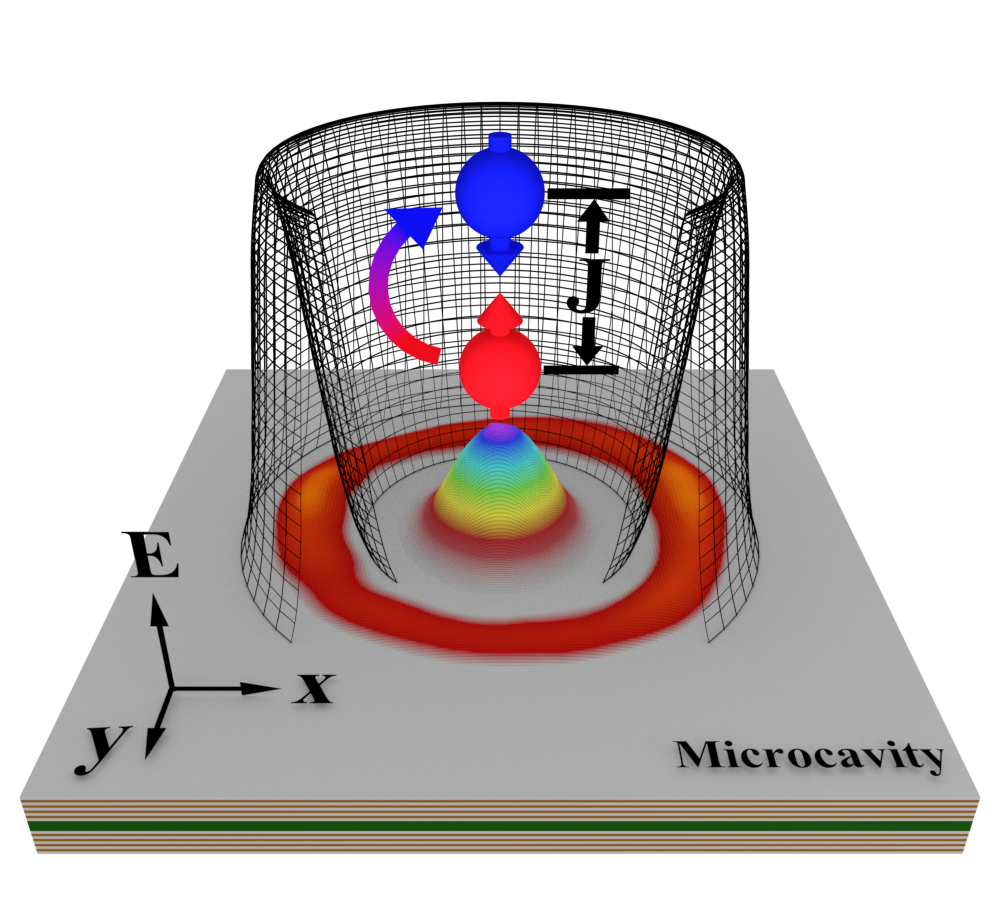}
\vspace{-0.75cm}
	\caption{Schematic of the optical trap regime employed. The red ring on the surface of the microcavity is an intensity map of the pump beam, the black mesh annular barrier represents the confining potential due to the exciton reservoir. A section of the barrier is removed to show the trapped condensate shown as the rainbow (near) Gaussian mode in the center of the confining potential and the red and blue spheres show the spin-up and spin-down states forming the condensate respectively. The arrow between the two spin states shows a current of polaritons from the spin-up to the spin-down state.}
	\label{F1}
	\vspace{-0.75cm}
\end{figure}

In this letter, we realize spin-bistability under non-resonant optical pumping, in the absence of external fields, by spatially separating the spin-bistable polariton state from the hot-exciton reservoir, thus strongly suppressing spin depolarisation in the system. The spin-bistability is evidenced on the spinor of a polariton condensate, trapped at the centre of an annular excitation beam, in the form of a hysteresis loop in the degree of circular polarization versus the optical excitation power. We attribute the spin-bistability to an interplay of polariton nonlinearities with the internal Josephson coupling of the two spin components. Figure \ref{F1} depicts a schematic of the system; the red ring, on the surface of the microcavity, shows the pumping intensity profile, while the rainbow colored Gaussian, in the centre of the ring, shows the measured polariton density. The black mesh annular barrier represents the repulsive potential due to the optically injected hot-exciton reservoir. The red and blue spheres in the centre of the ring represent the two spin components of the polariton condensate that are coupled via an internal Josephson coupling.

\begin{figure}[t!]
	\center
	\includegraphics[width=8.6cm]{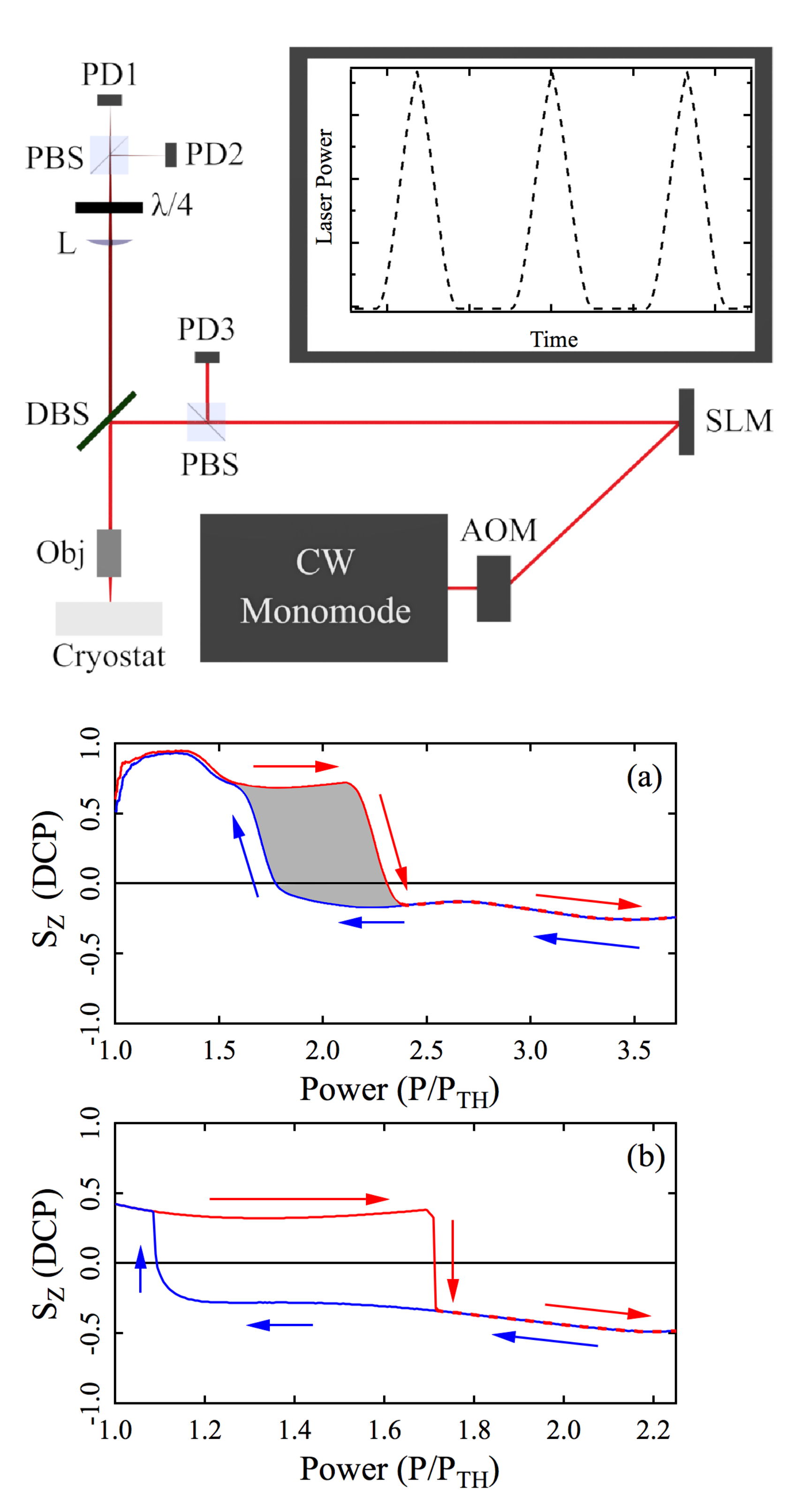}
\vspace{-0.75cm}
	\caption{Schematic of the experimental setup; SLM - spatial light modulator, AOM - acousto-optical modulator, PBS -polarizing beam splitter, DBS - dichroic beam splitter, Obj - microscope objective lens, L - plano-convex lens, $\lambda/4$ - quater waveplate and PD (1,2,3) photodiodes measuring the spin components of the signal and the corresponding laser intensity. (a) Measured third Stokes component, $S_{Z}$, representing the degree of circular polarization (DCP) vs laser power displaying the characteristic hysteresis of bistability. (b) Hysteresis loop in $S_{Z}$ vs laser power resulting from the numerical simulation of the Ginsburg-Landau based model.}
	\label{F2}
	\vspace{-0.75cm}
\end{figure}

The microcavity we use in this study is a $5\lambda/2$ cavity with GaAs quantum wells (QWs), as in \cite{kammann_nonlinear_2012}, held at $\sim$6 K using a continuous flow cold finger cryostat. We excite the sample with a linearly polarized, continuous wave (CW), monomode laser, tuned to the first reflectivity minimum at 754 nm. A spatial light modulator (SLM) imprints a phase profile such that the desired annular pump geometry is focused on the sample surface by a 0.4 numerical aperture (NA) microscope objective lens. The intensity of the laser is modulated into triangular pulses using an acousto-optical modulator (AOM) that is driven by a triangular voltage pulse train. A polarization analyser is used to record the two cross-circular components of the photoluminescence intensity. The top schematic of Figure \ref{F2} shows the setup and the triangular laser pulse train used to excite the sample. The use of a high NA excitation lens introduces a $\sim$10$\%$ ellipticity to the pump beam \cite{SpinWhirlsPRB}, which creates a small spin imbalance of the injected carriers. Since the electron spin relaxation time in GaAs QWs is longer than the carrier relaxation time to the exciton reservoir, the initial electron spin imbalance is reflected in the spin imbalance of the exciton reservoir that in turn leads to a spin polarized polariton condensate \cite{Alexis_Nonres}.

Figure \ref{F2}(a) shows a typical hysteresis loop in the degree of circular polarization (DCP), represented by the $S_Z$ component of the Stokes polarization vector versus the pump power normalized to the condensation threshold of the dominant (spin-up here) component. With increasing pump power, shown with a red line in Fig.2(a), we observe a spin reversal as it was previously observed in time integrated measurements \cite{Alexis_Nonres}.  This spin reversal was described as the transition into a desynchronization regime in Ref.\cite{Alexis_Nonres} through the complex spin-dependent Ginsburg-Landau equations (GLE) with an internal Josephson coupling term between the two spinor components. With decreasing pump power (blue line), a backwards spin reversal is observed at a lower pumping power compared to the power for the spin reversal observed in the forward direction. The signature of spin bistability is qualitatively reproduced numerically with the same theoretical model of Ref.\cite{Alexis_Nonres} as shown in Fig. {\ref{F2}(b); see appendix A of the Supplemental Material.

To gain understanding of the nature of polariton spin bistability, in the following we draw analogues with the driven damped pendulum. We non-dimensionalize the complex spin-dependent Ginsburg-Landau equation and the rate equation describing the density in the exciton-reservoir and reduce them to the equation of a driven-damped pendulum:
\begin{equation}
\ddot{\Theta} + \beta (p)  \dot{\Theta} =  -I_{\rm bias}(p) - I_{\rm cr}(p) \sin \Theta,
\label{Driven_Dumped_Pendulum}	
\end{equation}
where $\Theta$ is the phase difference between  two spin components, $\beta$ is a damping coefficient (positive above threshold) that increases with pumping power approaching a value of $1/2$, $I_{\rm bias}$ is equivalent to the driving torque of a pendulum and $I_{\rm cr}$ is equivalent to the maximum gravitational torque. A description of the process of non-dimensionalization and reduction is presented in appendix A of the Supplemental Material.

A driven damped pendulum supports two types of solution \cite{strogatz_nonlinear_2000}: if the torques' ratio $I = |I_{\rm bias}|/|I_{\rm cr}| \leq 1$, then the pendulum's trajectory is attracted to a fixed point (stationary solutions regime), where the angular displacement is constant ($\dot{\Theta}= 0 $) for a fixed driving torque. Alternatively, if $I>1$, then $\dot{\Theta}\neq 0 $, in which case the pendulum continues to overturn even under a fixed driving torque (limit cycle solutions regime). Thus bistability occurs in the backward ramp of the driving torque due to its inertia; the backward ramp corresponds to decreasing pump power here.

\begin{figure}[t!]
	\center
	\includegraphics[width=8.6cm]{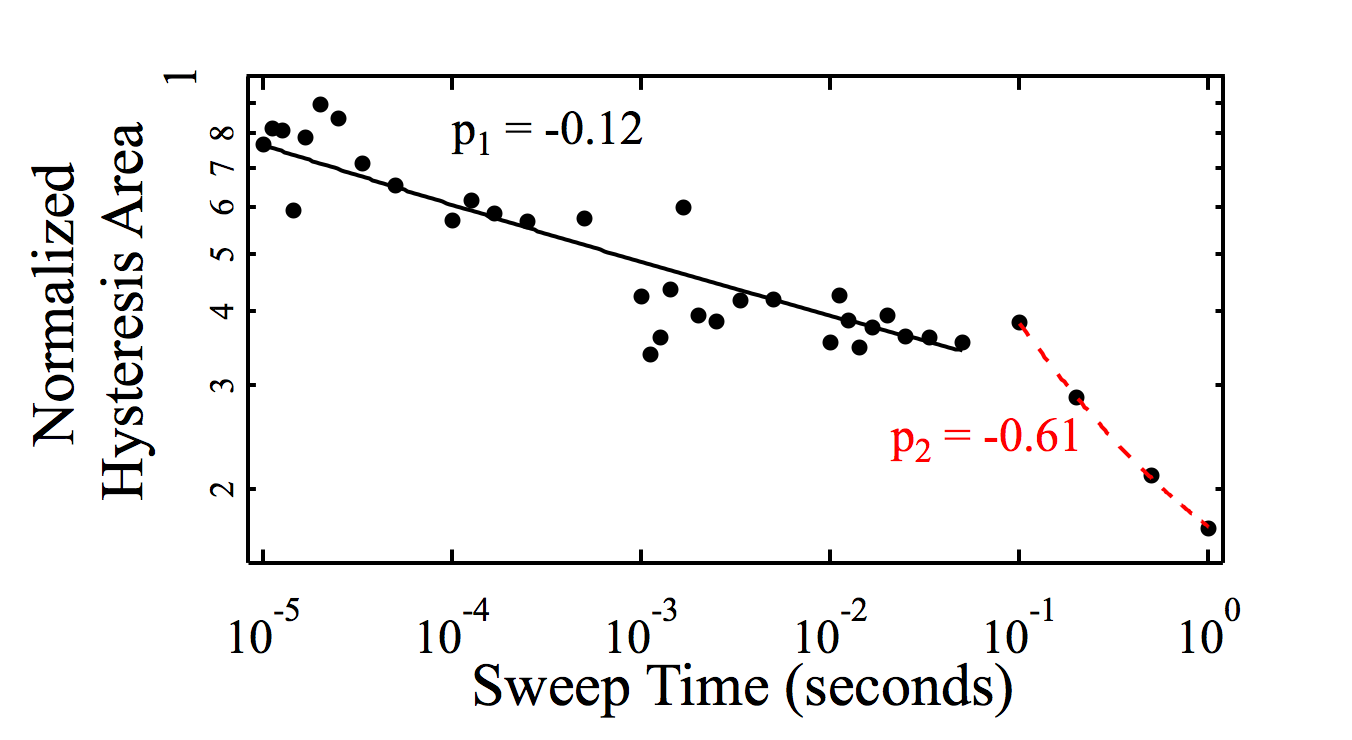}
	\centering
    \vspace{-0.75cm}
	\caption{Hysteresis area as a function of the sweep time, $T_{P}$, shown with black dots. The black (red) solid (dashed) line is a power-law fit for the fast (slow) sweep times with an exponent of -0.12 (-0.61). }
	\label{F3}
\vspace{-0.75cm}
\end{figure}

In the case of the spinor polariton condensate, the stationary solutions regime results in the phases of the two spin states evolving synchronously with a constant phase difference; top branch of the bistability in Fig. \ref{F2}(a,b). The limit cycle solutions regime, $\dot{\Theta}\neq 0 $, results in the desynchronised evolution of the spin states' phases. With decreasing pump power, the role of the ``inertia" increases due to the decrease of the damping rate $\beta$, which allows for the limit cycle solutions regime to persist for pump powers lower than those on the forward ramp, leading to the observed hysteresis. A schematic comparison between bistability in the driven damped pendulum and the spinor polariton condensate here, is presented in appendix B of the Supplemental Material.

In the following we investigate the hysteresis area on the duration of the sweep time ($T_{P}$). Figure \ref{F3} shows the hysteresis area for sweep times  spanning five orders of magnitude $T_{P} \in [10 \mu sec,1  sec]$ \cite{RatioNote}. We observe that the dependence of the hysteresis area as a function of the sweep time exhibits a double power-law decay. The mean-field approach of the spin-dependent Ginzburg-Landau model implemented here to explain the observed bistability cannot reproduce the dependence of the hysteresis area on the sweep time. Recently, a double power-law decay vs the sweep time was attributed to quantum fluctuations \cite{CiutiBistability_Theory}. An evolution from a double to a single power-law decay was experimentally observed with increasing the average photon number and was ascribed to a dissipative phase transition between the quantum regime and the thermodynamic limit \cite{rodriguez2016dynamic}. The dissipative phase transition was shown to depend strongly on the laser-cavity detuning. Here, we explore a new regime, wherein spinor bistability is observed in polariton condensates under non-resonant optical pumping.  Although, the exponents of the power-law fits follow the same trend as in the resonant excitation experiments, -bigger exponent for slower sweep times-, the values of the exponents are very different, probably due to the non-resonant pumping conditions.

In summary, we demonstrate optical bistability in a non-resonantly optically pumped spinor polariton condensate in the absence of external biasing fields. We model the occurrence of spin bistability through the complex spin-dependent Ginsburg-Landau equations (GLE) with an internal Josephson coupling term between the two spinor components. Furthermore, we observe a characteristic double power-law dependence of the hysteresis area versus the sweep time that cannot be reproduced within the context of mean-field theory. The observed dynamics of the hysteresis suggest that bistability in spinor polariton condensates under non-resonant pumping can serve as a test-bed for investigating the Kibble-Zurek mechanism and dynamic critical phenomena \cite{RevModPhys.49.435}.

\vspace{-0.5cm}
\section{Acknowledgements}
\vspace{-0.5cm}
The authors acknowledge the support of the Skoltech NGP Program (Skoltech-MIT joint project), and the UK’s Engineering and Physical Sciences Research Council (grant
EP/M025330/1 on Hybrid Polaritonics).

\widetext
\setcounter{equation}{0}
\setcounter{figure}{0}
\setcounter{table}{0}
\setcounter{page}{1}
\renewcommand{\theequation}{S\arabic{equation}}
\renewcommand{\thefigure}{S\arabic{figure}}
\renewcommand{\bibnumfmt}[1]{[S#1]}
\renewcommand{\citenumfont}[1]{S#1}

\pagebreak
\section{Optical bistability under non-resonant excitation in spinor polariton condensates: Supplemental Material}

\section*{Appendix A. Numerical Simulations.}

The complex spin-dependent Ginsburg-Landau model coupled to the reservoir rate equation is shown in Equations \ref{Eq1} and \ref{Eq2}.

\begin{eqnarray}
	i\hbar \frac{\partial \psi_{\pm}(t)}{\partial t} &=& \big\{\alpha_0 |\psi_{\pm}|^2 +  \alpha_1 |\psi_{\mp}|^2 + \hbar g_R n_{\pm} + \nonumber \\
	&+&  \frac{i \hbar}{2} \left(R_R n_{\pm}   - \gamma_C \right) \big\} \psi_{\pm} + \Omega \psi_{\mp},
	\label{Eq1}
\end{eqnarray}

\begin{equation}
	\frac{\partial n_{\pm}(t)}{\partial t}  = - \left( \gamma_R + R_R |\psi_{\pm}|^2 \right) n_{\pm} (t) + P_{\pm} (t),
	\label{Eq2}
\end{equation}	

Where  $\alpha_0>0$ is the strength of the repulsive interaction between polaritons with the same spin, $\alpha_1<0$ characterizes a weak attractive interaction between polaritons of opposite spin and $\Omega$ is the Josephson coupling term. $\gamma_{R}$ and $\gamma_{C}$ are the decay rates of the reservoir and condensate respectively, $R_{R}$ is the scattering rate from the reservoir into the polariton condensate, $P_{\pm}$ is the relative pumping rate of the reservoirs for each spin state and $g_{R}$ is the blueshift originating from interparticle interactions.

To understand the nature of the bistability we non-dimentionalize Eqs. (\ref{Eq1}-\ref{Eq2})  using $\psi_\pm \rightarrow  \sqrt{\hbar \gamma_C / \alpha_0} \psi_\pm, n_\pm \rightarrow 2 m \gamma_C / \hbar  n_\pm, t \rightarrow t/\gamma_{C}$   and introduce $U_\alpha=1 - \alpha_1/\alpha_0$, $g = 2 m g_R/\hbar, R = 2 m R_R/\hbar,  J = \Omega/\hbar \gamma_C,
\gamma = \gamma_R/\gamma_C,  b = \hbar R_R/\alpha_{0},  p_{\pm} = \hbar/2m \gamma_C^2 P_{\pm}, \epsilon = \eta - 1, \rho=(|\psi_+|^2+|\psi_-|^2)/2.$ In the "Josephson regime" where $J \ll U_{\alpha} \rho$, $ S_{Z}\ll 1$, and with a sufficiently small pumping imbalance $\epsilon \ll 1$, Eqs. (\ref{Eq1}-\ref{Eq2})  reduce to the equation of a driven-damped pendulum
\begin{equation}
	\ddot{\Theta} + \beta (p)  \dot{\Theta} =  -I_{\rm bias}(p) - I_{\rm cr}(p) \sin \Theta,
	\label{Driven_Dumped_Pendulum_SupInfo}
\end{equation}
where $\Theta$ is equal to the phase difference between the two spin components, $\beta (p) =  1/2 - \gamma/2 R p (1 + \epsilon),$ $I_{\rm bias}(p) = U_{\alpha}\rho_{\rm st} \epsilon/2 (2 + \epsilon),$ $I_{\rm cr}(p) = J \rho_{\rm st} (U_{\alpha} - g b/(1+ \epsilon) R^2 p)$ and $\rho_{\rm st} = [(\epsilon + 2)Rp - 2 \gamma]/2b$ is the stationary solution (a Bloch surface) of $d\rho/dt=0.$ Figure \ref{S3} shows the results of the simulation regarding the dependence of $S_{Z}$ on the pumping intensity, normalized to the threshold of condensation $P_{TH}$. The dimensionless parameters used in the simulation are $\gamma = 9$, $U_{\alpha} = 1.08$, $g = 4.4$, $b = 4$, $\eta = 1.12$, $J = 0.45$.

\begin{figure}[b!]
	\centering
	\includegraphics[width=8.6cm]{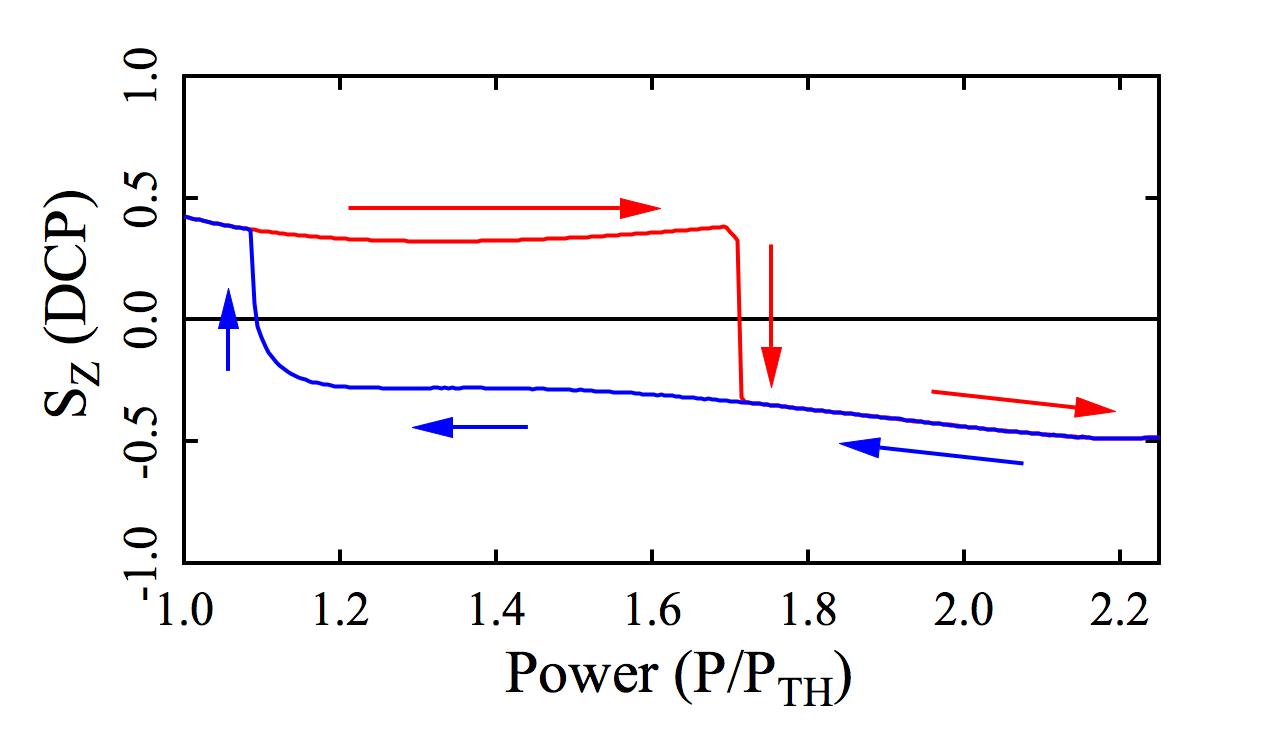}
	\caption{The numerical results from modeling the dynamics of Equations (\ref{Eq1}-\ref{Eq2}) showing the dependence of the degree of circular polarization (DCP), which is represented by the $S_{Z}$ component of the Stokes vector, on the pumping intensity as it is swept forward (up) and then backward (down) in a symmetrical triangular fashion. }
	\label{S3}
\end{figure}

Starting with random initial conditions (ICs) within the bistability region, where $S_{Z}$ is double valued, it is possible to get solutions in either the stationary solutions regime (synchronized) or limit cycle solutions regime (desynchronized). Using the coordinates $\rho$ and $S_{Z}$ we display how the solutions evolve across the simulation time domain in Figure \ref{S4}(a) for $P/P_{TH}=1.2$. Figure \ref{S4}(b) shows the end point of the said evolution and demonstrates the coexistence of the limit cycle solutions regime and the stationary solutions regime. By considering a pumping power closer to the switching power we observe that the previous solution in the stationary solutions regime transforms into a limit cycle solution (Figure \ref{S5}); which then converge as $S_{Z}$ switches.

Starting again within the bistability region we consider which parameters are critical in determining whether a stationary solutions regime or limit cycles solutions regime is reached. It is found that the phase difference $\Theta$ and degree of circular polarization ($S_{Z}$) are significant. The resulting bifurcation diagram for a pump power $P/P_{TH}=1.2$ is displayed in Fig \ref{S4}(c) and its shape is valid for any pump power within the bistability region.

\begin{figure}[t!]
	\centering
	\includegraphics[width=9cm]{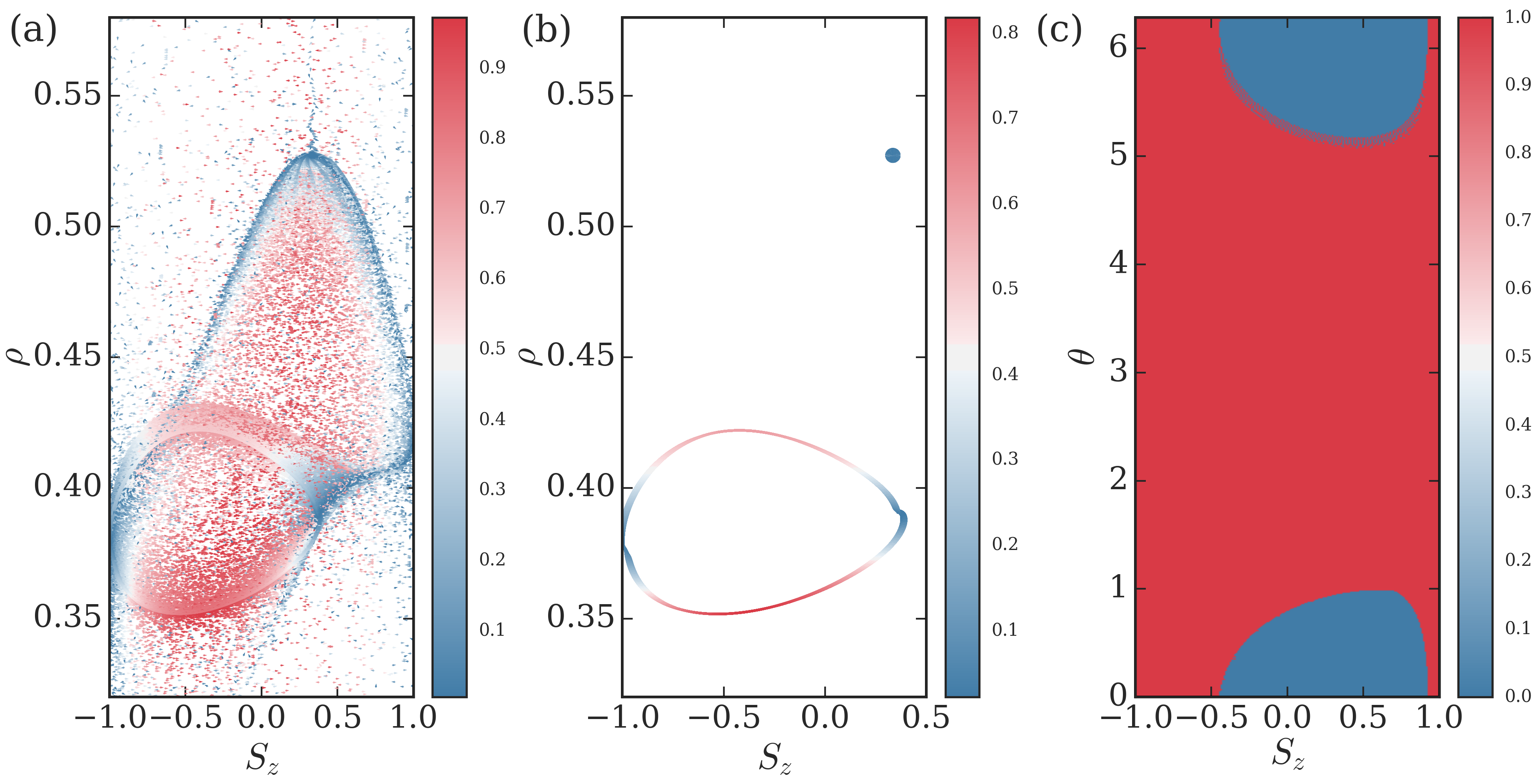}
	\caption{The characteristic shape shows how starting from the random IC the solution evolves in time (a) and we approach either fixed point or limit cycle (b) (red corresponds to high velocities). (c) The bifurcation diagram in terms of the dependence of the phase difference between the species $\Theta$ and the DCP $S_z$: starting from the points marked with red the system ends in a limit cycle solution while the blue points lead to fixed point solutions. All the plots are simulated for $P/P_{th} = 1.2$.}
		\label{S4}
\end{figure}

\begin{figure}[!h]
	\centering
	\includegraphics[width=9cm]{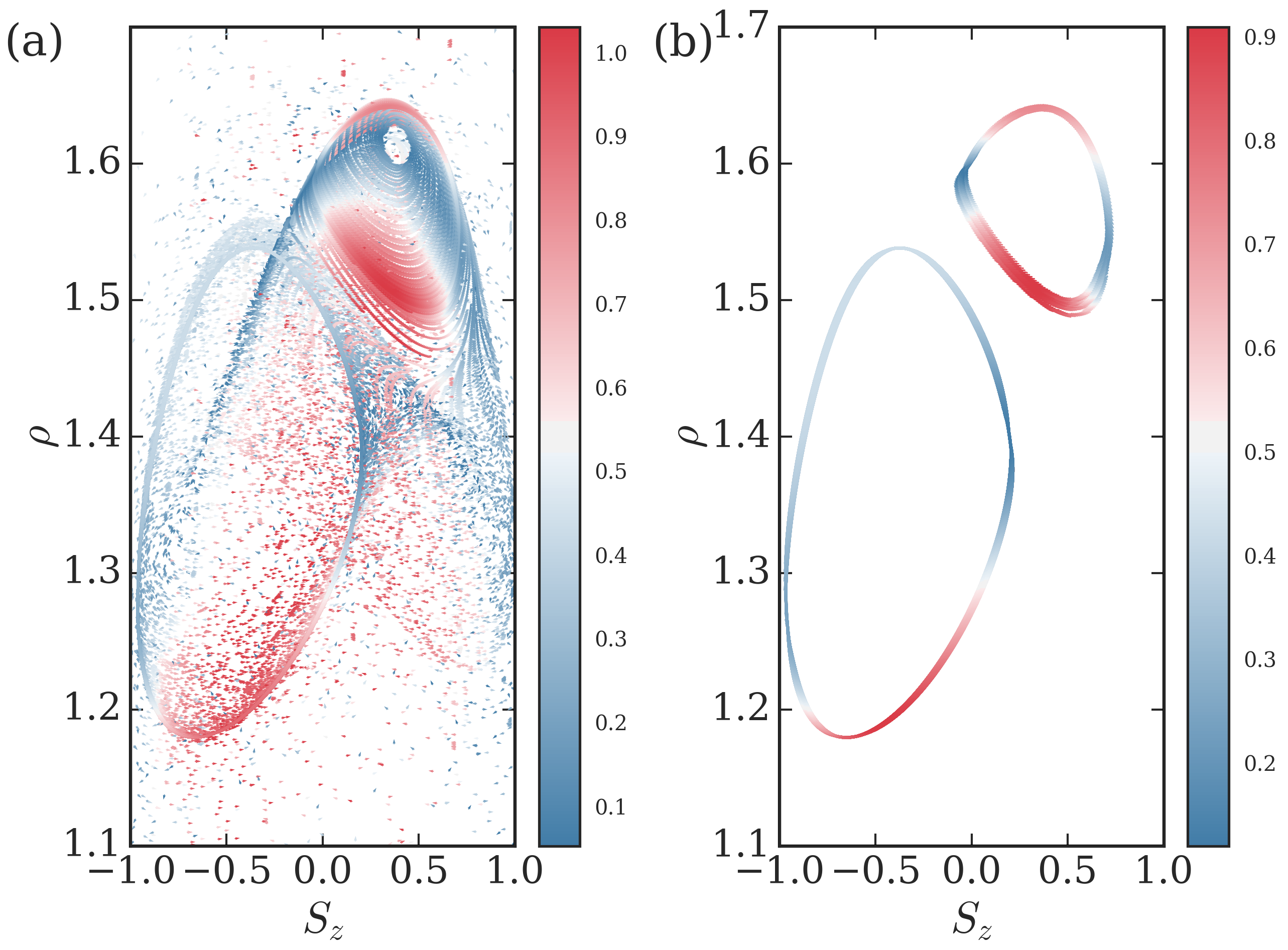}
	\caption{Phase portrait for a pump power nearing the transition point in which $S_{Z}$ changes sign. (a) Shows the path taken during the running time of the simulation. (b) Depicts the double limit cycle solution the simulation converges on. }
		\label{S5}
\end{figure}

\section*{Appendix B. Pendulum Analogy.}

A system analogous to the non-resonantly pumped spinor polariton condensate is a driven damped pendulum; comparisons between the two are shown schematically in Figure \ref{SM_B1}. As discussed in the main text, driven damped pendula can exist in two regimes, the stationary solutions regime and the limit cycle solutions regime. The stationary solutions regime exists when the pendulum maintains a constant angular displacement ($\dot{\Theta} = 0 $) from the vertically down position under constant driving torque. The limit cycle solutions regime corresponds to when the pendulum can overturn ($\dot{\Theta}\neq 0 $) even with a constant driving torque. For the purpose of comparison between the driven damped pendulum and our non-resonantly pumped spinor polariton condensate the stationary solution regime is referred to as the synchronized regime wherein the phase difference between spin components remains constant ($\dot{\Theta} = 0 $). The limit cycle solutions regime is referred to as the desynchronized regime wherein $\dot{\Theta}\neq 0 $.

On the forward sweep of the driving torque, the pendulum starts in the synchronized regime and remains there until the pendulum reaches the point of maximum gravitational torque (horizontal position); point 2 in Figure \ref{SM_B1}. As the driving torque is increased beyond this point the pendulum transitions from the synchronized to the desynchronized regime, which corresponds to the change in sign of $S_{Z}$ in the polariton condensate. On the backward sweep of the driving torque the pendulum is initially overturning, in the desynchronized regime, then as the driving torque is reduced below that corresponding to point 2 in Figure \ref{SM_B1} the overturning of the pendulum persists due to the inertia (in this case angular momentum). When the driving torque is reduced sufficiently that the pendulum is no longer able to rotate, in the same direction, the pendulum transitions back to the synchronized regime closing the hysteresis loop.

\begin{figure*}
	\center
	\includegraphics[width=17.2cm]{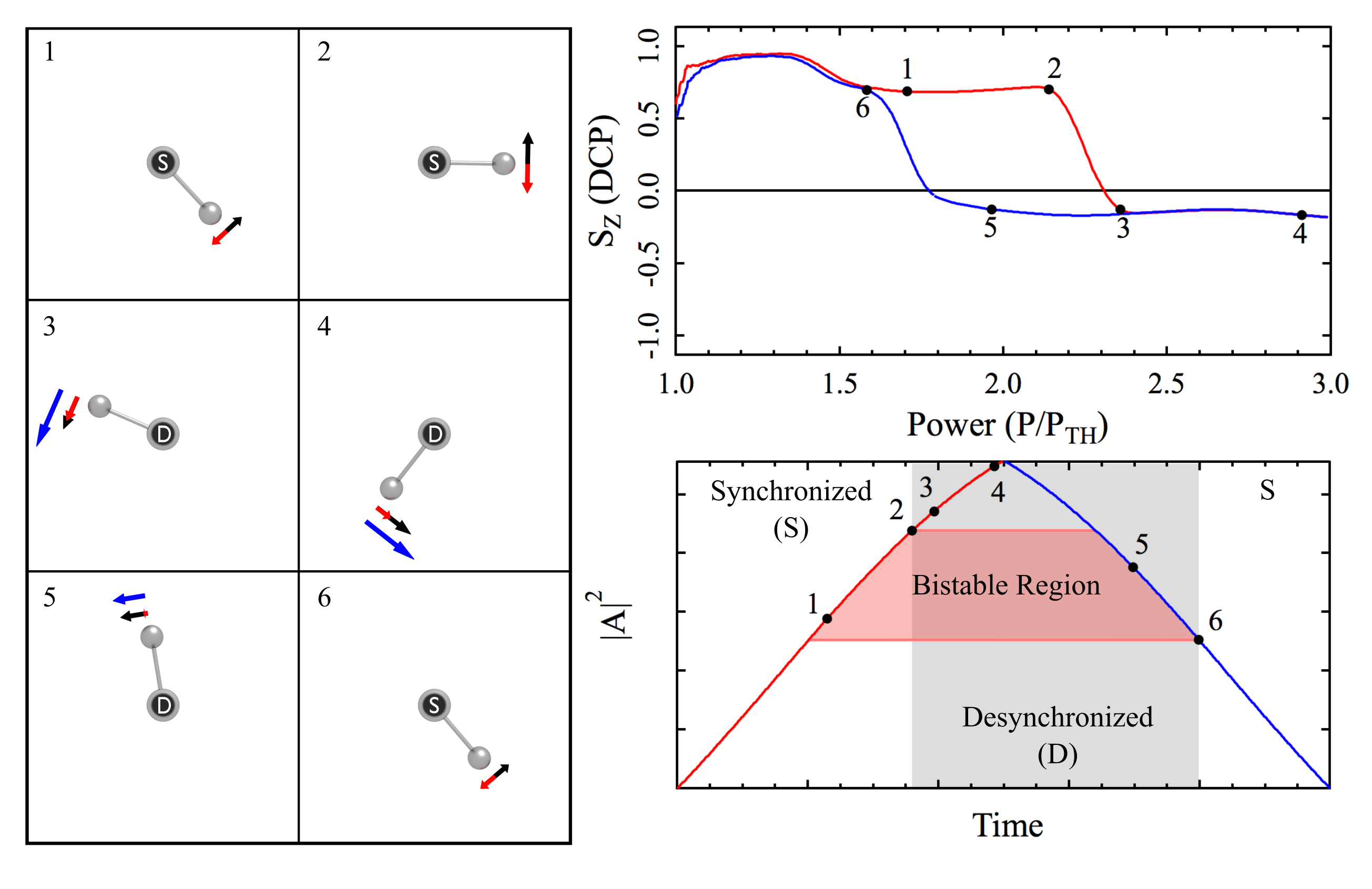}
	\centering
	\caption{1-6 Schematic of the state of a driven damped pendulum  with arrows representing the torque; the driving torque (black), the gravitational torque (red) and the total torque (blue). The positions corresponding to 1-6 are highlighted on the plots of $S_{Z}$ vs pump power and $|A|^{2}$ vs time, where $|A|^{2}$ is the pump intensity in the case of the spinor polariton condensate and the magnitude of the driving torque in the case of the driven damped pendulum.}
	\label{SM_B1}
\end{figure*}


\begin{thebibliography}{00}
\bibitem{RatioNote} For all the measurements here, the ratio of the time interval between successive sweeps, $T_{Sep}$, and the sweep time, $T_{P}$, is kept constant at $T_{Sep}/T_{P}=5$.
\end{thebibliography}
\end{document}